\documentclass[
  a4paper, %
  10pt, %
  twoside, %
]{LTJournalArticle}

\usepackage[british]{babel}
\runninghead{Surface tension-driven boundary growth in tumour spheroids} %

\footertext{Accepted for publication in \textit{Interface Focus} --- Themed issue in \textit{Capillarity and Elastocapillarity in Biology}} %

\title{Surface tension-driven boundary growth\\ in tumour spheroids} %

\author{%
  D. Riccobelli$^{1,\,2}\,$\thanks{email: \href{mailto:davide.riccobelli@sissa.it}{davide.riccobelli@sissa.it}}
}

\date{\footnotesize $^1$Mathematics Area, mathLab, SISSA, via Bonomea 265, 34136, Trieste, Italy\\
$^2$MOX -- Dipartimento di Matematica, Politecnico di Milano, Piazza Leonardo da Vinci 32, 20133, Milano, Italy}

\renewcommand{\maketitlehookd}{%
  \begin{abstract}
    \noindent Growing experimental evidence highlights the relevant role of mechanics in the physiology of solid tumours, even in their early stages. While most of the mathematical models describe tumour growth as a volumetric increase of mass in the bulk, \emph{in vitro} experiments on tumour spheroids have demonstrated that cell proliferation occurs in a thin layer at the boundary of the cellular aggregate. In this work, we investigate how elasticity and surface tension interact during the development of tumour spheroids. We model the spheroid as a hyperelastic material undergoing boundary accretion, where the newly created cells are deformed by the action of surface tension. This growth leads to a frustrated reference configuration, resulting in the appearance of residual stress.
    Our theoretical framework is validated using experimental results from the literature. Like fully developed tumours, spheroids open when subjected to radial cuts. Remarkably, this behaviour is observed even in newly formed spheroids, which lack residual stress. Through both analytical solutions and numerical simulations, we show that this phenomenon is driven by elastocapillary interactions, where the residual stress developed in grown spheroids amplifies the tumour opening. Our model’s outcomes align with experimental observations and allow us to estimate the surface tension acting on tumour spheroids.
  \end{abstract}
}

\AtEveryBibitem{\clearfield{month}}
\AtEveryCitekey{\clearfield{month}}
\usepackage{amsthm, amsmath,dblfloatfix}
\theoremstyle{plain}

\DeclareMathOperator{\Diver}{Div}
\DeclareMathOperator{\Grad}{Grad}

\DeclareMathOperator{\tr}{tr}

\usepackage{mathtools}
\usepackage{pgfplots}
\pgfplotsset{/pgf/number format/use comma,compat=newest}

\renewcommand\epsilon{\varepsilon}
\newcommand{\R}{\mathbb{R}}

\newcommand{\vect}[1]{\boldsymbol{#1}}
\newcommand{\tens}[1]{\mathbf{#1}}

\renewcommand{\d}{\mathrm{d}}
\newenvironment{system}[1]{\left\{
  \begin{aligned} #1}{
  \end{aligned}\right.}

\begin{document}

\maketitle %

\section{Introduction}
Uncontrolled cell proliferation is the main feature that distinguishes healthy tissues from tumours: while the physiological growth of tissue is tightly regulated and self-limited by cell signalling, tumour cells replicate in an unregulated manner. The huge social impact of cancer has given the impulse to extensive research focused on understanding the mechanisms underlying tumour growth.

Cancer research frequently relies on surrogates to investigate the dynamics of tumour growth and the effects of drugs, ranging from experiments performed \emph{in vivo} to \emph{in vitro}. In the first type of experiments, tumours are implanted and grown in animals, such as mice \cite{Stylianopoulos_2012}. Conversely, \emph{in vitro} experiments involve the cultivation of tumour cells outside a living organism in the laboratory~\cite{Zanoni_2020}. A notable example is provided by multicellular tumour spheroids, which are three-dimensional structures that self-assemble, forming spherical aggregates.

Tumour spheroids serve as an artificial model mimicking the early stages of tumour development. At this stage, the cells within the spheroid rely on passive intercellular diffusion for the supply of nutrients and oxygen, as blood vessels have not yet penetrated the tumour mass.

Research on tumour spheroids has revealed a strong connection between mechanical forces and tumour proliferation \cite{Jain_2014}: applying compressive load at the boundary reduces cell mitosis and can induce apoptosis \cite{Helmlinger_1997,Montel_2011,Montel_2012,Mascheroni_2016,Ambrosi_2017,Dolega_2021,Erlich_2023}. Furthermore, mechanical pressure has been linked to increased invasiveness in cancer cells \cite{Pandey_2024}.

\begin{figure*}[t!]
  \centering
  \includegraphics[width=\columnwidth]{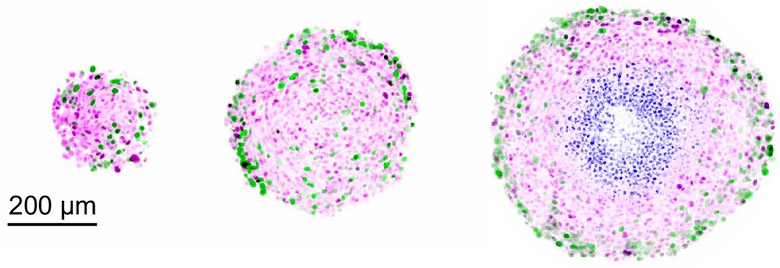}\qquad\vrule\qquad
  \includegraphics[width=0.4\columnwidth]{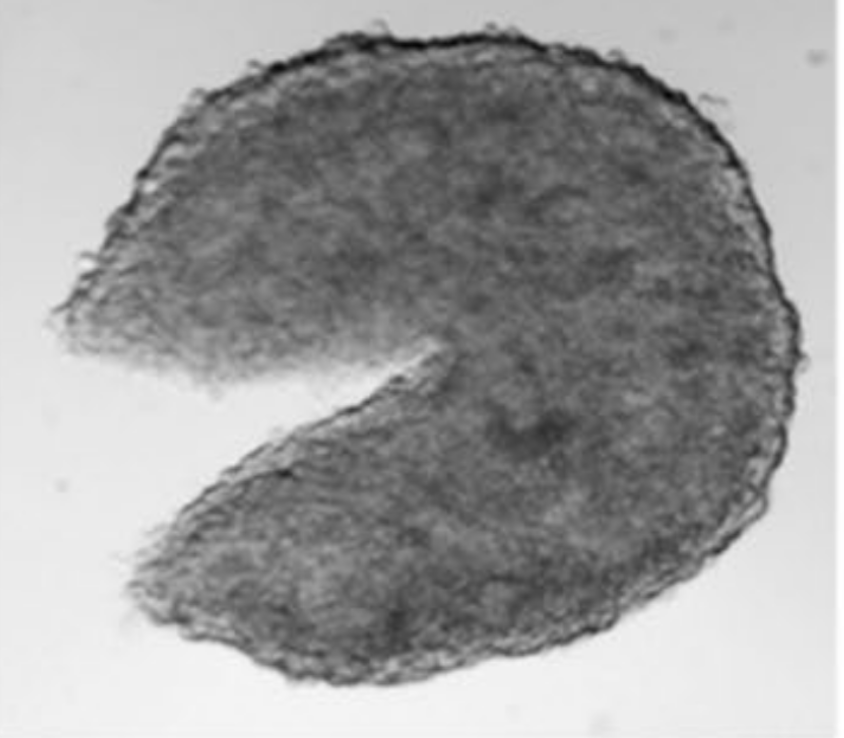}
  \caption{(Left) Sections of tumour spheroids with varying radii. Recently dividing cells are coloured green, while necrotic cells are indicated in blue (adapted from \cite{Browning_2021}). (Right) An incised tumour spheroid two days post-seeding, with an initial cell count of 5000 (adapted from \cite{Guillaume_2019}).}
  \label{fig:browning}
\end{figure*}

In order to understand how boundary loads drive the inner mechanical state, a rheological description of spheroids is essential. Many authors describe spheroids and cellular aggregates as fluids subject to surface tension \cite{Davis_1997,Foty_1994}. However, solid matter can also exhibit surface tension effects, especially when surface forces dominate over bulk elasticity. This interplay between elasticity and surface tension has given rise to the field of \emph{elastocapillarity} \cite{style2017elastocapillarity,Bico_2018}.

An increasing body of evidence indicates that cellular aggregates exhibit viscoelastic behaviour, with elasticity playing a fundamental role. The mechanical response is strongly influenced by surface tension, particularly in aggregates with small characteristic lengths, such as tumour spheroids. For instance, elasticity and surface tension govern the formation of brain sulci during brain organoid growth \cite{Riccobelli_2020}. Similarly, in \cite{Yadav_2022} the authors analyse the spatial reorganization of tumour cells when a spheroid is locally exposed to laser ablation, inducing a surface tension gradient at the spheroid boundary. Their findings reveal that the resulting cell motion differs from classical Marangoni flow in fluids, underscoring the importance of the spheroid's elastic properties in the process. Furthermore, in \cite{Ang_2024} the authors investigate elastocapillary effects in glioblastoma cell spheroids embedded in a collagen matrix. Their study demonstrates that matrix contraction and deformation are driven by the interplay of spheroid surface tension, cell-collagen interactions, and poroelastic effects.

Interestingly, similarly to fully developed tumours, spheroids tend to open when subjected to a radial cut \cite{Colin_2018,Guillaume_2019,Kosheleva_2023}. If the spheroid obeyed a fluid-like behaviour governed by surface tension, it would reorganize to minimize surface area, returning to a spherical shape. However, experimental evidence shows that this is not the case.
Indeed, radially cut tumour spheroids display a heart-like shape, see Fig.~\ref{fig:browning}. Moreover, even intact tumour spheroids can exhibit both spherical and cylindrical shapes depending on the initial number of cells \cite{Guillaume_2019}, suggesting that while surface tension plays a role in the spheroid physics, other forces also influence its shape.

The mechanical origin of the deformation following the cut is a matter of debate. In \cite{Colin_2018,Guillaume_2019}, the authors argue that the spheroid opening is driven by non-homogeneous volumetric growth within the tumour. In biological systems, non-homogeneous growth can lead to geometrical incompatibilities that generate mechanical stress, known as \emph{residual stress} \cite{Rodriguez_1994}, even in the absence of external load. Volumetric growth and residual stresses have been extensively studied in recent years \cite{Mascheroni_2016,Ambrosi_2017,Dolega_2021}, leading to mathematical models that explain a variety of phenomena in biological matter. This has given rise to a field of research known as \emph{morphoelasticity} \cite{Goriely_2006,Ambrosi_2017,Goriely_2017}. Successful applications of morphoelasticity include modelling organ and tissue morphogenesis \cite{Taber_1995,taber2000modeling,Balbi_2013,Ben_Amar_2013}, understanding deformations in arteries \cite{Chuong_1986,Holzapfel_2009}, and explaining mechanical instabilities in tumour vessels~\cite{Araujo_2004,MacLaurin_2012,Riccobelli_2018}.

However, several aspects remain obscure. For instance, the residual stress responsible for tumour opening following a radial cut is peculiar, requiring tensile hoop residual stress in the outer region of the spheroid \cite{Jain_2014,Ambrosi_2017}. Such a stress state can be produced only if cell proliferation decreases near the free surface \cite{Walker_2023}. The situation in reality is the opposite: even in the early stages of spheroid development, cell mitosis occurs in a narrow region close to the spheroid boundary \cite{Montel_2011,Montel_2012,Delarue_2013,Browning_2021}, see Fig.~\ref{fig:browning}. Moreover, the self-aggregation process of tumour cells leading to the formation of tumour spheroids usually takes around two days. In this timeframe, no residual stress should have formed, and thus no opening would be expected when a spheroid is cut. However, Guillaume et al. \cite{Guillaume_2019} observed spheroid openings even in these early stages.

During the initial days, only a thin outer layer of cells experiences circumferential tension, while the core remains in an isotropic stress state \cite{Lee_2019}. This tensional ``skin'' at the spheroid's boundary can be understood as an effect of tissue surface tension on the cell aggregate \cite{Riccobelli_2020}. This surface phenomenon arises from the combined effects of passive intercellular cohesion (mediated by E-cadherin) and active cortical contractility (driven by myosin II) \cite{manning2010coaction}, with evidence suggesting a greater contribution from the latter \cite{Yadav_2022,Yousafzai_2022}. The coupling of elasticity and surface tension is therefore a key determinant of the mechanical behaviour of cellular aggregates, particularly at the length scales characteristic of multicellular spheroids \cite{Riccobelli_2020,Yadav_2022,Yousafzai_2022,Kosheleva_2023,Ang_2024}.

The aim of this work is to construct a mathematical model of tumour spheroid growth able to explain the tumour opening following a radial cut, as reported in the experiments of Guillaume et al. \cite{Guillaume_2019}. In Section~\ref{sec:model}, we propose the mathematical model of the spheroid, where tumour proliferation is described as a surface accretion mediated by surface tension. The study of spheroid incision is performed in Section~\ref{sec:incision}, where we propose both a simplified analytical prediction of tumour opening and a quantitative analysis through numerical simulations and comparison with experimental results. Finally, the main results are summarized in Section~\ref{sec:conclusions}, together with some concluding remarks.

\section{Surface tension and elasticity in tumour spheroids}
\label{sec:model}

In this section we construct a mathematical model of a tumour spheroid as an elastic body subject to surface tension. First, we describe the mathematical setting in the absence of cell proliferation. Second, we extend the theoretical framework to include surface growth mediated by capillarity effects.

\subsection{A continuum model of tumour spheroid with surface tension}

The tumour spheroid is treated as an elastic continuum body subject to surface tension. By following the theory of continuum mechanics, we introduce the reference configuration $\Omega_0\subset\mathbb{R}^3$. Let $(\vect{e}_X,\,\vect{e}_Y,\,\vect{e}_Z)$ be the canonical vector basis.

Let $\vect{\varphi}:\Omega_0\rightarrow\mathbb{R}^3$ be the deformation field, we denote by $\tens{F} = \Grad\vect{\varphi}$ the deformation gradient.
The current configuration of the body is identified by $\Omega$, i.e. $\Omega = \vect{\varphi}(\Omega_0)$.

Given a point $\vect{X}\in\Omega_0$, we denote by $\vect{x}$ the corresponding point in the current configuration, so that $\vect{x}=\vect{\varphi}(\vect{X})$. We can introduce the displacement field $\vect{u}$ as the vector function $\vect{u}(\vect{X}) = \vect{\varphi}(\vect{X})-\vect{X}$.

In order to take into account the large elastic deformations of the tumour, we assume that the cellular aggregate can be described as a hyperelastic material. Although tumour spheroids can accumulate extracellular matrix in their inner regions due to the formation of a necrotic core \cite{nederman1984demonstration}, the spheroids used in the experiments by Guillaume et al. \cite{Guillaume_2019} are relatively young and lack a necrotic core. Thus, it is reasonable to assume that the spheroid is homogeneous. Let $\psi_r = \psi_r(\tens{F})$ be the strain energy density. The \emph{bulk elastic energy} of the body is thus given by
\begin{equation}
  \label{eq:elastic_energy}
  \mathcal{E}_\text{el}[\vect{\varphi}] = \int_{\Omega_0} \psi_r(\tens{F})\,\d V.
\end{equation}

We can also introduce a \emph{surface energy} that models the action of tissue surface tension. As commonly done in the context of cellular aggregate, we take it as directly proportional with the current area $a$ of the external boundary $\sigma_\gamma$, i.e.
\begin{equation}
  \label{eq:surface_energy}
  \mathcal{E}_\text{s}[\vect{\varphi}] = \gamma a =  \gamma \int_{\Sigma_\gamma}\left(\det\tens{F} \right)|\tens{F}^{-T}\vect{N}| \d A,
\end{equation}
where $\gamma$ is the surface tension, $\vect{N}$ is the external normal in the reference configuration, $\Sigma_\gamma$ is the referential counterpart of the external boundary $\sigma_\gamma$, i.e. $\sigma_\gamma=\vect{\varphi}(\Sigma_\gamma)$.
The deformation $\vect{\varphi}$ must minimize the energy of the system $\mathcal{E}$, given by
\begin{equation}
  \label{eq:total_energy}
  \mathcal{E}[\vect{\varphi}] = \mathcal{E}_\text{el}[\vect{\varphi}] + \mathcal{E}_\text{s}[\vect{\varphi}].
\end{equation}

Cellular aggregates are nearly incompressible due to their high content of water. Here, we approximate their behaviour as incompressible, i.e. the deformation must obey the constraint
\begin{equation}
  \det\tens{F} = 1.
\end{equation}
The stress can be measured in the reference configuration by means of the first Piola--Kirchhoff stress tensor, given by
\begin{equation}
  \tens{S} = \frac{\partial\psi_r}{\partial\tens{F}} - p\tens{F}^{-T},
\end{equation}
where $p$ is a scalar field that imposes the incompressibility constraint.

The balance equations corresponding to the minimization of the energy functional \eqref{eq:total_energy} are given by
\begin{equation}
  \label{eq:nonlinearpb}
  \begin{system}
    & \Diver\tens{S} = \vect{0}                                 &  & \text{in }\Omega_0,      \\
    & \tens{S}\vect{N} = \gamma\mathcal{K}\tens{F}^{-T}\vect{N} &  & \text{on }\Sigma_\gamma, \\
  \end{system}
\end{equation}
where $\mathcal{K}$ is twice the mean curvature of the current surface $\sigma_\gamma$. Rigid body motions can be filtered out, for example, by fixing the displacement of a point of the body and by requiring zero mean rotation.

In order to proceed with the analysis, we have to make some constitutive choice on the mechanical behaviour of the spheroid. In the following, we assume that the body $\Omega_0$ is composed of a neo--Hookean material, so that the strain energy density reads
\begin{equation}
  \label{eq:strain_energy_density}
  \psi_r(\tens{F})=\frac{\mu}{2}\tens{F}:\tens{F},
\end{equation}
where $\mu$ is the shear modulus, while $:$ denotes the Frobenius scalar product of tensors, i.e. $\tens{A}:\tens{B} = \tr(\tens{A}^T\tens{B})$.

\subsection{Surface growth}
While the model proposed in the previous section can describe the elastic behaviour of a spheroid, it is insufficient to capture what occurs as the spheroid grows. Indeed, spheroids left to grow for several days in an environment with large availability of nutrients can generate mechanical stress as a consequence of their growth. The increase in the opening length of cut spheroids grown for a few days \cite{Guillaume_2019} suggests that the growing process enhances the natural opening induced by surface tension by generating a tensile hoop stress in the outer part of the spheroid.

Analogously to spheroids, incised real tumours open up, even though surface tension is negligible due to their larger radius \cite{Stylianopoulos_2012}. This deformation results from the release of the tensile hoop stress in the outer region, while there is a compressive radial residual stress in the core \cite{Ambrosi_2017}. Puzzlingly, to reproduce this stress pattern using classical volumetric growth theory, the growth of the tumour spheroid should be greater in the inner part rather than the outer part \cite{Ambrosi_2017,Walker_2023}, which is precisely the opposite of what is observed in spheroids.
In \cite{Browning_2021}, the proliferation of cells in tumour spheroids seems to be mainly concentrated at the periphery within a layer a few cells thick, while cells in the bulk duplicate only rarely, see Fig.~\ref{fig:browning}. Similar behaviours have been reported in \cite{Montel_2011,Montel_2012,Delarue_2013}.

Differently from previous models \cite{Mascheroni_2016,Ambrosi_2017,Dolega_2021,Erlich_2023,Walker_2023}, here we describe spheroid's growth as a surface accretion, where new mass is added at the boundary rather than in the bulk. We start by considering a time-dependant reference configuration, i.e. $\Omega_0(t)$. In particular, we consider $t\in[0,\,T]$, where $T$ is the final time instant. Let $R_t$ be the radius of the (intact) tumour spheroid at time $t$, so that
\[
  \Omega_0(t) = \{\vect{X}\in\mathbb{R}^3\;|\;\|\vect{X}\| < R_t\}.
\]
We indicate with $\vect{\varphi}_t$ the deformation field at time $t$, so that $\Omega(t)=\vect{\varphi}_t(\Omega_0(t))$ is the current configuration at a given instant of time. We denote by $(R,\,\Theta,\,\Phi)$ and $(r,\,\theta,\,\phi)$ the spherical coordinates in the reference and current configurations, respectively. Similarly, let $(\vect{e}_R,\,\vect{e}_\Theta,\,\vect{e}_\Phi)$ and $(\vect{e}_r,\,\vect{e}_\theta,\,\vect{e}_\phi)$ the corresponding vector basis.

As the spheroid grows, new particles are added to the system at the boundary of the domain. Due to the presence of surface tension, the spheroid is subject to a mechanical load at the boundary. From \eqref{eq:nonlinearpb}, we get
\begin{equation}
  \label{eq:Cauchy_ST}
  \tens{T}(\vect{x},\,\tens{F})\vect{n} = \gamma\mathcal{K}\vect{n},
\end{equation}
where $\tens{T} = \tens{S}\tens{F}^T$ is the Cauchy stress tensor. As the process of growth is much slower than the characteristic timescale of the deformation, we can safely neglect inertia and the balance equation \eqref{eq:nonlinearpb} still holds, with $\Sigma_\gamma$ corresponding to the whole boundary $\partial\Omega_0(t)$.

Among the several frameworks proposed to model surface growth, see for instance \cite{Skalak_1982,dicarlo2005surface,Ciarletta_2013,sozio2017nonlinear,von2021morphogenesis}, we follow the theory proposed by Truskinovsky and Zurlo in a series of works \cite{Zurlo_2017,Zurlo_2018,Truskinovsky_2019}. This approach accounts for the elastic distortions induced by the boundary condition \eqref{eq:Cauchy_ST} in the material generated on the surface.

\begin{figure}[t!]
  \includegraphics[width=\columnwidth]{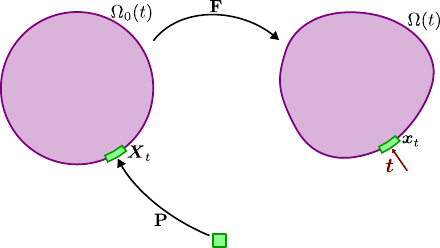}
  \caption{Representation of the material transplant of a point particle $\vect{X}_t$ \cite{Epstein_2012}. When a new point is placed at the boundary of the spheroid, it is first subject to an elastic distortion $\tens{P}$ from the relaxed state to satisfy the boundary condition $\tens{T}\vect{n}=\vect{t}$ after the deformation $\tens{F}$ is applied.}
  \label{fig:placement}
\end{figure}

To model the elastic frustration in the spheroid, we assume that $\psi_r$, introduced in \eqref{eq:strain_energy_density}, represents the strain energy density of the spheroid in its relaxed state for a volume element of mass density $\rho_0$. We call this function \emph{material archetype} \cite{Epstein_2012}. Specifically, during the deposition of a new particle $\vect{X}_t$ belonging to the boundary of $\Omega_0(t)$, the material is elastically distorted and the strain energy density in the reference configuration can be written as
\[
  \psi(\vect{X}_t,\,\tens{F}) = \frac{1}{\det\tens{P}(\vect{X}_t)}\psi_r(\tens{F}\tens{P}(\vect{X}_t)).
\]
We refer to $\tens{P}$ as the transplant map, or more simply as \emph{transplant}. Such a tensor models the local elastic distortion from the relaxed archetype to the reference configuration at point $\vect{X}_t$, see Fig.~\ref{fig:placement}.

While being on the boundary at time $t$, the point $\vect{X}_t$ belongs to the bulk of $\Omega_0(\bar{t})$ for $\bar{t}>t$. Once the material is deposed, here we assume that $\tens{P}$ does not change in time. The transplant $\tens{P}$ can be also extended to the initial reference configuration $\Omega_0(0)$ by taking $\tens{P}(\vect{X})=\tens{I}$ for $\vect{X}\in\Omega_0(0)$.
In this way, $\tens{P}$ is defined in the whole the reference domain for all $t$, and the energy functional becomes
\[
  \mathcal{E}_\text{el}[\vect{\varphi}] = \int_{\Omega_0} \psi(\vect{X},\,\tens{F})\,dV.
\]
Such an operation is called \emph{material transplant}. Generally speaking, the reference configuration may not be stress-free if the transplant map $\tens{P}(\vect{X})$ is not a gradient of a deformation function $\vect{\varphi}_\tens{P}(\vect{X})$, for a detailed discussion see \cite{Epstein_2012,Goriely_2017}.
In the following, the explicit dependence of $\tens{P}$ on $\vect{X}$ is omitted unless necessary.

The total elastic distortion of the material can be computed as $\tens{F}_e=\tens{F}\tens{P}$, accounting for the total deformation from the archetype to the current configuration.
Within this framework,  the first Piola--Kirchhoff stress tensor reads
\begin{equation}
  \label{eq:Piola}
  \tens{S} = \frac{\partial\psi}{\partial\tens{F}}-p\tens{F}^{-T}=\frac{1}{\det\tens{P}}\frac{\partial\psi_r}{\partial\tens{F}_e}\tens{P}^T-p\tens{F}^{-T}.
\end{equation}

\begin{figure*}[ht!]
  \includegraphics[width=\textwidth]{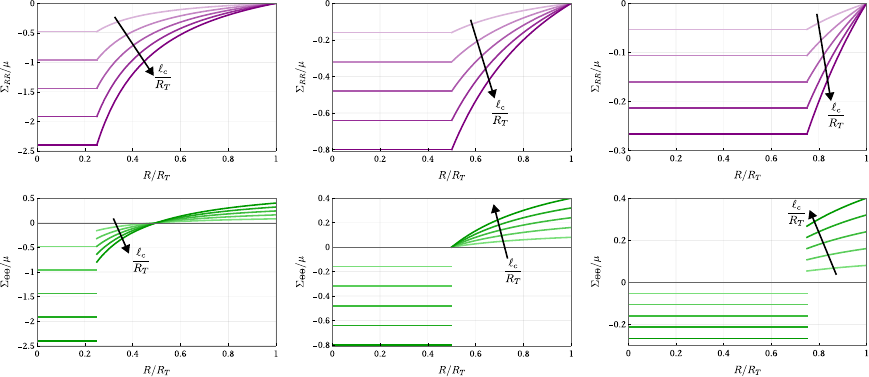}
  \caption{Plots of dimensionless radial (top) and hoop (bottom) residual stresses, normalized by the shear modulus $\mu$, as a function of the radial position $R/R_T$. The panels correspond to $R_0/R_T$ values of 0.25 (left), 0.5 (centre), and 0.75 (right). Each curve represents different values of the normalized capillary length $\ell_c/R_T = 0.04, 0.08, \dots, 0.2$, where $\ell_c = \gamma/\mu$. The direction of increasing $\ell_c/R_T$ is indicated by arrows, with darker curves representing higher values of $\ell_c/R_T$.}
  \label{fig:res_stress}
\end{figure*}
During the process of deposition of the new cells on the boundary at $\vect{X}_t$, the stress state of the newly deposed material must satisfy the boundary condition \eqref{eq:Cauchy_ST}. Specifically, we take
\begin{equation}
  \label{eq:deposition}
  \tens{T}(\vect{x}_t,\,\tens{F}) = \gamma\mathcal{K}\vect{n}\otimes\vect{n},
\end{equation}
where $\vect{x}_t = \vect{\varphi}_t(\vect{X}_t)$.%
\footnote{
  According to Zurlo and Truskinovsky~\cite{Zurlo_2017}, at the time of the deposition
  \[
    \tens{T}(t,\,\vect{x}_t,\,\tens{F}) = \tens{T}_p + \tens{T}_e,
  \]
  where $\tens{T}_p$ is exactly the right-hand side of \eqref{eq:deposition} and satisfies the boundary conditions, while $\tens{T}_e$ is a given tensor field accounting for the stress state in other directions, i.e. with $\tens{T}_e\vect{n}=\vect{0}$. In the following we assume that there is no force other than the action of the boundary condition \eqref{eq:Cauchy_ST}, and we take $\tens{T}_e =\tens{0}$.
}
Finally, provided that the density of the added point is $\rho_0$, i.e. the same as the archetype, we get \cite{Epstein_2012}
\begin{equation}
  \label{eq:incP}
  \det \tens{P} = 1.
\end{equation}

In order to find the explicit expression of the transplant, we observe that before cutting the spheroid has a spherical shape. Therefore, we can assume that $\tens{P}$ is a function of the radial coordinate only, with
\[
  \tens{P}(R) = P_{RR}(R)\vect{e}_R\otimes\vect{e}_R+P_{\Theta\Theta}(R)(\tens{I}-\vect{e}_R\otimes\vect{e}_R),
\]
where $P_{RR}$ and $P_{\Theta\Theta}$ are (unknown) components of the transplant map to be determined.
From \eqref{eq:incP}, we immediately get $P_{\Theta\Theta} = P_{RR}^{-1/2}$.
We observe that $\vect{\varphi}_t(\vect{X}) =\vect{X}$ is a solution of \eqref{eq:nonlinearpb}. Indeed, the incompressibility constraint $\det\tens{F} = 1$ is satisfied and the first Piola--Kirchhoff stress tensor coincides with the Cauchy stress. By taking $\tens{F} = \tens{I}$ in \eqref{eq:Piola} we get
\begin{equation}
  \label{eq:stress_no_def}
  \tens{S} = \tens{T} = \mu \tens{P}\tens{P}^T - p\tens{I}
\end{equation}
By substituting \eqref{eq:stress_no_def} into \eqref{eq:deposition} and observing that $\mathcal{K}=-2/R_t$, after a few passages we are left with the following third-order equation
\begin{equation}
  \label{eq:PRR_eq}
  \mu  \left(P_{RR}^3(R_t)-1\right) R_t+2 \gamma  P_{RR}(R_t)=0.
\end{equation}
The only positive real solution of such an equation is given by $P_{RR}(R_t) = F(R_t)$, where
\[
  \begin{gathered}
    F(R_t)=\frac{\sqrt[3]{2} \left(G(R_t)\right)^{2/3}-4 \sqrt[3]{3} \ell_c R_t}{6^{2/3} R_t \sqrt[3]{G(R_t)}},\\
    G(R_t)=\sqrt{96 \ell_c^3 R_t^3+81 R_t^6}+9 R_t^3,
  \end{gathered}
\]
where $\ell_c$ is the elastocapillary length, defined as
\begin{equation}
  \label{eq:capillary_length}
  \ell_\text{c} = \frac{\gamma}{\mu}.
\end{equation}
Summarizing, at $t=T$ we have
\[
  P_{RR}(R) =
  \begin{system}
    & F(R) &  & \text{it } R\geq R_0, \\
    & 1    &  & \text{if }R<R_0.
  \end{system}
\]

Finally, the balance equation \eqref{eq:nonlinearpb} is always satisfied thanks to the pressure variable $p$. Specifically, under spherical symmetry conditions \eqref{eq:nonlinearpb} reduces to
\begin{equation}
  \label{eq:balance_equation_spherical}
  \frac{\d S_{rR}}{\d R} + \frac{2 (S_{rR}-S_{\theta\Theta}) }{R}=0.
\end{equation}
By substitution of \eqref{eq:stress_no_def} in \eqref{eq:balance_equation_spherical}, we can integrate with respect to the radial coordinate from $R_t$ to $R$. Recalling the boundary condition \eqref{eq:Cauchy_ST}, we get
\[
  p(t, R) =
  \begin{system}
    & \mu P_{RR}(R)^2+2 \gamma  \left(\frac{2}{R}-\frac{1}{R_t}\right)
    &                                                                  & \text{if }R\geq R_0, \\
    & \mu\left(1-P_{RR}(R_0)^2\right)+p(t,\,R_0)
    &                                                                  & \text{if }R<R_0.
  \end{system}
\]

The distribution of the transplant map is spatially inhomogeneous. This can give rise to a stress state even if the surface tension is removed at the outer boundary after the growth of the spheroid. Such a stress state is called \emph{residual stress}, and it is due by the geometrical incompatibilities induced by surface growth \cite{Zurlo_2017}, similarly to what happens with volumetric growth \cite{Rodriguez_1994,Goriely_2017}.

Since there is no deformation induced by surface tension, the residual stress tensor $\tens{\Sigma}$ can be simply computed as
\[
  \tens{\Sigma} = \mu \tens{P}\tens{P}^T - p_\tens{\Sigma},
\]
where $p_\tens{\Sigma}$ is the pressure field in the absence of surface tension. From the boundary condition $\tens{\Sigma}\vect{N}=\vect{0}$ we get
\[
  p_\tens{\Sigma}(R)=p(T,\,R)+\frac{2\gamma}{R_T}.
\]
Therefore, the radial and hoop components of the residual stress read
\[
  \Sigma_{RR}=
  \begin{system}
    & 4 \gamma  \left(\frac{1}{R_T}-\frac{1}{R}\right),   &  & \text{if }R\geq R_0, \\
    & 4 \gamma  \left(\frac{1}{R_T}-\frac{1}{R_0}\right), &  & \text{if }R< R_0,
  \end{system}
\]\[
  \Sigma_{\Theta\Theta}=
  \begin{system}
    & 2 \gamma\left(\frac{2 }{R_T}-\frac{1 }{R}\right),   &  & \text{if }R\geq R_0, \\
    & 4 \gamma  \left(\frac{1}{R_T}-\frac{1}{R_0}\right), &  & \text{if }R< R_0.
  \end{system}
\]

\begin{figure*}[!hb]
  \includegraphics[width=\textwidth]{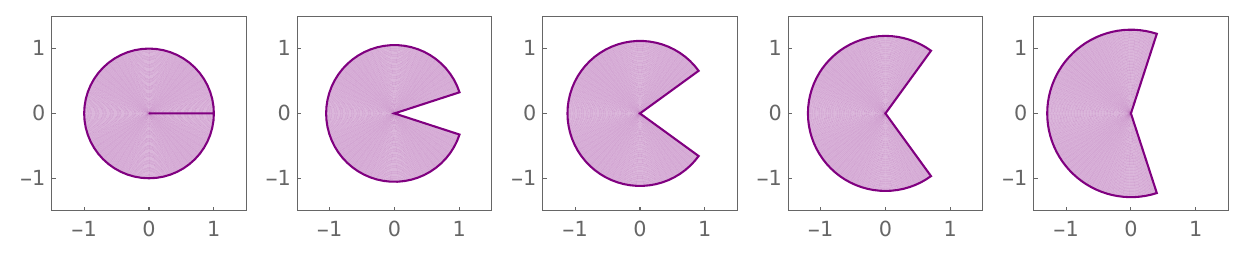}
  \caption{Representation of the ``Pac-Man'' deformation described by \eqref{eq:pacman} for $\alpha = 1, 0.9,\,\dots,\,0.6$ from left to right. All the lengths are rescaled with respect to the initial radius $R_0$. The deformation is applied to the whole disk for the sake of clarity.}
  \label{fig:pacman}
\end{figure*}

We observe that the resulting residual stress is independent of the material properties, but its magnitude is proportional to the surface tension $\gamma$. Representative residual stress profiles are shown in Fig.~\ref{fig:res_stress}. From these plots, it is clear that the radial residual stress vanishes at $R=R_T$ (consistent with the boundary conditions for residual stress), remains continuous at $R=R_0$, and is compressive (negative) throughout the spheroid. In contrast, the hoop residual stress is tensile in the outer region, becomes zero at $R = R_T/2$ (if $R_0 < R_T/2$), and is compressive in the inner region.
In any case, the magnitude of the hoop residual stress at the boundary is independent on $R_0$ and depends only on the ratio $\ell_c/R_T$. Both the radial and the hoop stress are equal in the centre $R<R_0$, in agreement with several experimental results \cite{Dolega_2017,Lee_2019}

Thus, the model returned the desired pattern of residual stress, with a tensile region in the outer rim and a compressive stress state in the inner region. In the next section, we analyse the impact of surface tension and growth on the opening of the spheroid due to a radial cut.

\section{Deformation of a radially cut tumour spheroid}
\label{sec:incision}

In this section, we analyse the deformation of a tumour spheroid following a radial incision. The experiments described by \cite{Guillaume_2019} investigate the deformation of both newly formed cellular aggregates and more mature ones, with cuts applied at various stages of growth. Many of these experiments focus on aggregates consisting of approximately 5000 cancer cells, which initially exhibit a shape closer to a cylinder rather than a sphere. Importantly, the observed opening in newly formed spheroids upon cutting cannot be attributed to spatially inhomogeneous growth. As discussed previously, this conclusion is supported by experimental evidence showing that only a thin layer of cells at the boundary undergoes circumferential stretch after two days \cite{Lee_2019}.

We begin by presenting a simplified two-dimensional model to describe the cutting of a recently formed tumour spheroid, without considering surface growth. After this, we extend the analysis to a fully three-dimensional model that incorporates growth for spheroids that have been allowed to develop for several days.

\subsection{A two-dimensional model of early-stage spheroid cutting}

Our goal is to understand the effects of a radial incision on early-stage tumour spheroids. Differently from Section~\ref{sec:model}, we first consider a reference configuration where the spheroid is approximated as a cylinder, which is cut radially, with the cylinder's axis aligned along the $Z$-axis. To simplify the analysis, we assume that deformations along the $Z$-axis are negligible, allowing us to focus solely on deformations in the plane perpendicular to the axis.

To model the impact of a radial cut with zero thickness—where no material is removed and a sharp interface is created—we further simplify the problem by analysing only half of the cross-section, assuming axial symmetry throughout.

More specifically, let $\mathcal{S}_0$ be half of the section of the cylinder section lying on the $XY$ plane
\[
  \mathcal{S}_0 = \{(R\cos\Theta,\,R\sin\Theta)\in\R^2\},
\]
with $0< R < R_0$ and $0<\Theta<\pi$.
We assume that the portion of the boundary subject to surface tension $\Sigma_\gamma$ is given
\[
  \begin{small}
    \Sigma_\gamma = \{\vect{X} = (X,\,Y)\;|\; Y=0\,\cap\,X>R_0-L\;\text{or}\;|\vect{X}|=R_0\},
  \end{small}
\]
where $0<L<2R_0$ is the length of the incision.

We apply symmetry boundary condition on the boundary $\Sigma_s = \{\vect{X} = (X,\,Y)\;|\; Y=0\,\cap\,X<R_0-L\}$, namely
\begin{equation}
  \label{eq:disp_u_1}
  \begin{system}
    & \vect{u}\cdot\vect{e}_Y = 0,           \\
    & \tens{S}\vect{e}_Y\cdot\vect{e}_X = 0.
  \end{system}
\end{equation}
Finally, in order to avoid rigid displacements of the half-disk, we set
\begin{equation}
  \label{eq:disp_u_2}
  \vect{\varphi}(0,0) = \vect{0}.
\end{equation}

In the following, we provide an analytical estimate of the opening length induced by surface tension, using a simplified kinematic model.

\subsubsection{Analytical estimate using simplified kinematics}

We now show that, by restricting the set of the admissible deformation fields, we are able to provide an analytical estimate of the opening length of a disk subject to a cut of length $L=R_0$. The polar coordinates in the current configuration are $(r,\,\theta)$ %

We restrict the kinematics to a specific set of deformations: we assume that the actual polar coordinates are given by \cite{Singh_1965,Silling_1991}
\begin{equation}
  \label{eq:pacman}
  \begin{system}
    & r = \frac{R}{\sqrt{\alpha}},            \\
    & \theta = \alpha \Theta + (1-\alpha)\pi,
  \end{system}
\end{equation}
where $\alpha \in (0,1]$ is a parameter describing the tumour opening, with $\alpha = 1$ corresponding to the undeformed tumour and the tumour opening increasing as $\alpha$ decreases.

The transformation introduced in \eqref{eq:pacman} describes the deformation of a half-disk into a disk sector of the same area. By applying this deformation symmetrically to the entire disk, we obtain a ``Pac-Man''-like shape, as shown in Fig.~\ref{fig:pacman}.

The admissible deformation $\vect{\varphi}_\alpha$ defined in \eqref{eq:pacman} is quite restrictive, but it provides enough freedom for the system to release surface energy through elastic deformation, neglecting the rounding of the corners induced by surface tension \cite{Mora_2015}. We observe that for the admissible deformations
\begin{equation}
  \label{eq:def_grad_simple}
  \begin{aligned}
    \nabla\vect{\varphi}_\alpha & = \frac{\d r}{\d R}\vect{e}_r\otimes\vect{e}_R+\frac{r}{R}\frac{\d\theta}{\d\Theta}\vect{e}_\theta\otimes\vect{e}_\Theta = \\
    & = \frac{1}{\sqrt{\alpha}}\vect{e}_r\otimes\vect{e}_R+\sqrt{\alpha}\vect{e}_\theta\otimes\vect{e}_\Theta,
  \end{aligned}
\end{equation}
so that the deformation gradient has constant coefficients using a polar vector basis.

\begin{figure}[b!]
  \includegraphics[width=\columnwidth]{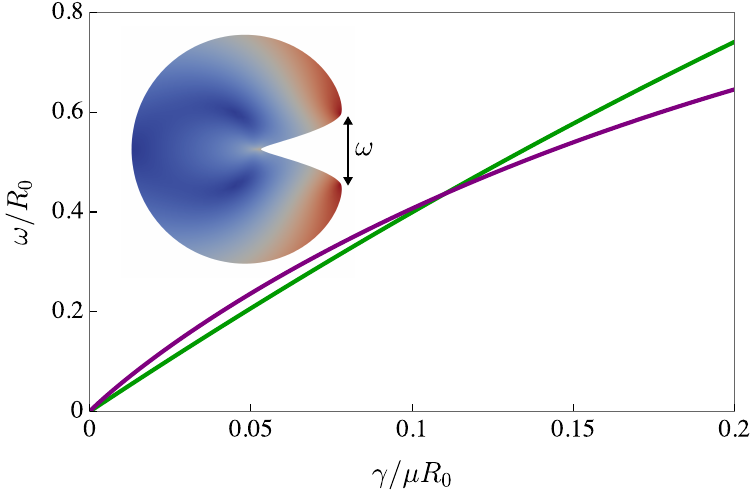}
  \caption{Opening length $\omega$ as a function of the surface tension $\gamma$ for a radial cut of length $R_0$ of a disk. The purple curve denote the numerical prediction, while the green curve is the result of the single-degree-of-freedom model of Eqs.~\eqref{eq:implicit_alpha}-\eqref{eq:omega}. All the quantities are non-dimensionalised with respect to the radius $R_0$ and the shear modulus $\mu$.}
  \label{fig:th_num_2D}
\end{figure}

Thanks to \eqref{eq:def_grad_simple}, the elastic energy stored in the body \eqref{eq:elastic_energy} can be written as
\begin{equation}
  \label{eq:simplified_el_energy}
  \begin{aligned}
    \mathcal{E}_\text{el}[\vect{\varphi}_\alpha] = \frac{\pi\mu R_0^2}{4}\left(\alpha + \frac{1}{\alpha}\right);
  \end{aligned}
\end{equation}
while the surface energy is given by
\begin{equation}
  \label{eq:simplified_surf_energy}
  \mathcal{E}_\text{s}[\vect{\varphi}_\alpha] = \gamma R_0\left(\frac{1}{\sqrt{\alpha}}+\pi\sqrt{\alpha}\right).
\end{equation}
The energy minima are stationary points of the functional $\mathcal{E}[\vect{\varphi}_\alpha]$, i.e.
\[
  \frac{\d}{\d\alpha}\mathcal{E}[\vect{\varphi}_\alpha]= \frac{\d}{\d\alpha}(\mathcal{E}_\text{el}[\vect{\varphi}_\alpha]+\mathcal{E}_\text{s}[\vect{\varphi}_\alpha]) =0,%
\]
which gives
\begin{equation}
  \label{eq:implicit_alpha}
  \frac{\ell_c}{R_0} =\frac{\pi  \left(1-\alpha ^2\right)}{2 \sqrt{\alpha } (\pi  \alpha -1)}.
\end{equation}
The expression given by \eqref{eq:implicit_alpha} can be inverted numerically to find  $\alpha$. %
Given the opening angle, it is possible to compute the opening length $\omega$ of the cut cylinder as
\begin{equation}
  \label{eq:omega}
  \omega  =2\vect{u}(0,\,R_0)\cdot\vect{e}_X= \frac{2 R_0\sin\alpha}{\sqrt{\alpha}}.
\end{equation}
A plot of the opening predicted by this single-degree-of-freedom model is reported in Fig.~\ref{fig:th_num_2D}, where we observe an almost linear relation between $\omega$ and the elastocapillary length $\ell_c$.

Of course, this is a very simplified model, where the energy depends only on a single scalar variable, that is $\alpha$. We are restricted to considering just a cut length equal to the radius and other effects. Moreover, the deformation close to sharp corners induced by surface tension is neglected \cite{Mora_2013,Mora_2015}.  In the following, we compare the proposed approximation by with numerical simulations.

\subsubsection{Finite element approximation}

The problem of spheroid cutting cannot be solved analytically if we allow non-homogeneous deformations. Therefore, we have to rely on numerical simulations. In what follows, we adopt the finite element method. Specifically, the problem can be written in weak form as
\begin{equation}
  \label{eq:finite_element_pb}
  \begin{gathered}
    \text{Find }(\vect{u},\,p)\text{ such that, for all }(\vect{v},\,q):\\
    \int_\mathcal{S}\tens{S}:\nabla\vect{v}\,\d V - \int_{\Sigma_\gamma}\gamma\mathcal{K}\vect{v}\cdot\tens{F}^{-T}\vect{N}\,\d S +\\
    - \int_\mathcal{S} q(\det\tens{F}-1)\,\d V = 0,
  \end{gathered}
\end{equation}
where $(\vect{v},\,q)$ are the test functions. Both $(\vect{u},\,p)$ and $(\vect{v},\,q)$ must belong to an appropriate functional space ensuring that the boundary conditions \eqref{eq:disp_u_1} and \eqref{eq:disp_u_2} are satisfied. We do not enter into the mathematical details, we refer to \cite{Bonet_2008,Quarteroni_2014} for a detailed discussion.

The half-disk domain, $\mathcal{S}$, is discretized using a triangular mesh. Since surface tension induces singular stresses near sharp angles \cite{Mora_2015}, the mesh is refined around the vertices of the cut to improve the accuracy of the numerical solution.
\begin{figure}[t!]
  \includegraphics[width=\columnwidth]{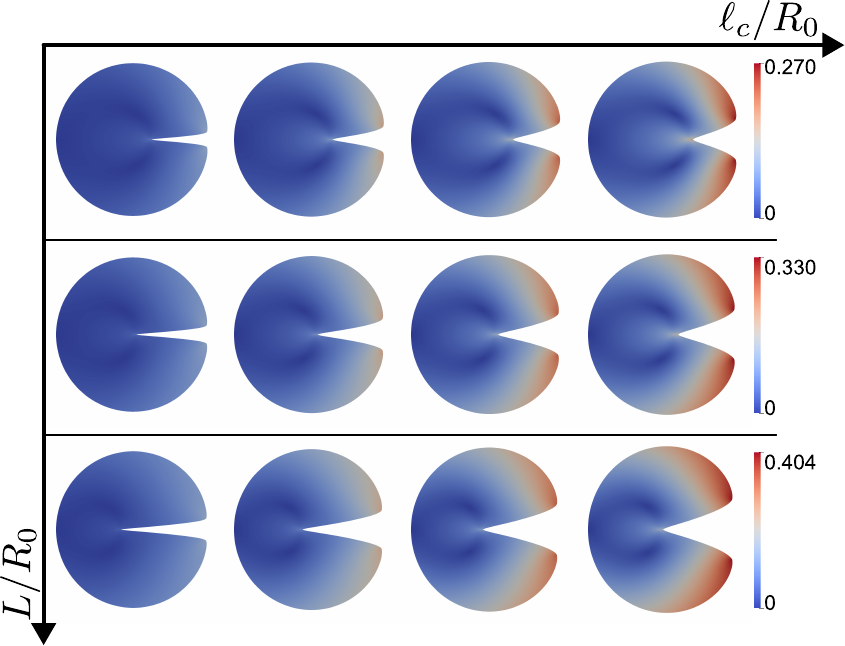}
  \caption{Morphological diagram showing the current configuration of the 2D model of the cut spheroid for $\ell_c/R_0 = 0.05,\,0.1,\,0.15,\,0.2$ and $L/R_0 = 0.8, 1, 1.2$. The colour bar denotes the norm of the non-dimensionalised displacement $\vect{u}/R_0$.}
  \label{fig:2D_map}
\end{figure}
\begin{figure}[t!]
  \includegraphics[width=\columnwidth]{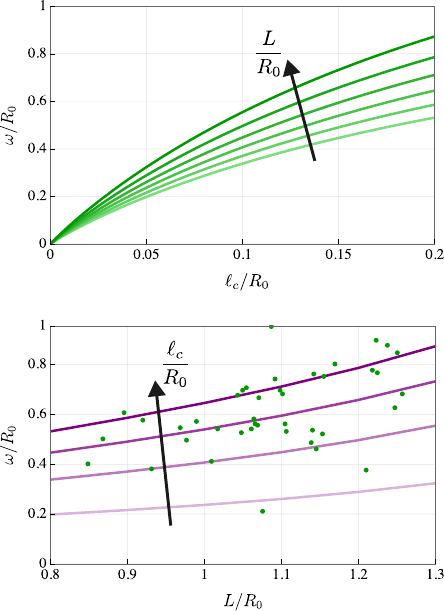}
  \caption{Plot of the opening $\omega$ of the cut disk as a function of the elastocapillary length (top) and of the incision length (bottom). All lengths are non-dimensionalised with respect to $R_0$. The value of $L/R_0$ in the top panel are $0.8,\,0.9,\,\dots,\,1.3$, while the values of $\ell_c/R_0$ in the bottom panel are $0.05,\,0.1,\,0.15,\,0.2$. In the bottom panel we also show the experimental data for a spheroid with $5000$ cells at the beginning of the experiment and cut after two days.}
  \label{fig:2D_data}
\end{figure}
We employ a mixed formulation with Taylor-Hood elements, discretizing the displacement and pressure fields using piecewise quadratic and linear functions, respectively. This choice ensures the well-posedness of the discrete problem and prevents the emergence of spurious pressure modes \cite{Boffi_2013}.

The implementation is carried out using the FEniCS library \cite{langtangen2017solving}. Surface tension is incrementally increased by a step size $\Delta \gamma$. The problem \eqref{eq:finite_element_pb} is solved using Newton's method, with the solution from $\gamma = \gamma_k$ serving as the initial guess for the subsequent Newton iteration at $\gamma = \gamma_{k+1} = \gamma_k + \Delta \gamma$. The increment $\Delta \gamma$ is automatically adjusted if the Newton method fails to converge or converges in fewer than three iterations. This algorithm is implemented using the library BiFEniCS \cite{Riccobelli_2021}.

\subsubsection{Results of the two-dimensional simulations}

In Fig.~\ref{fig:th_num_2D}, we compare the theoretical predictions of the single-degree-of-freedom analytical model given by \eqref{eq:implicit_alpha}-\eqref{eq:omega} with the outcomes of the numerical simulations (with an incision length of $L=R_0$).

\begin{figure}[b!]
  \includegraphics[width=\columnwidth]{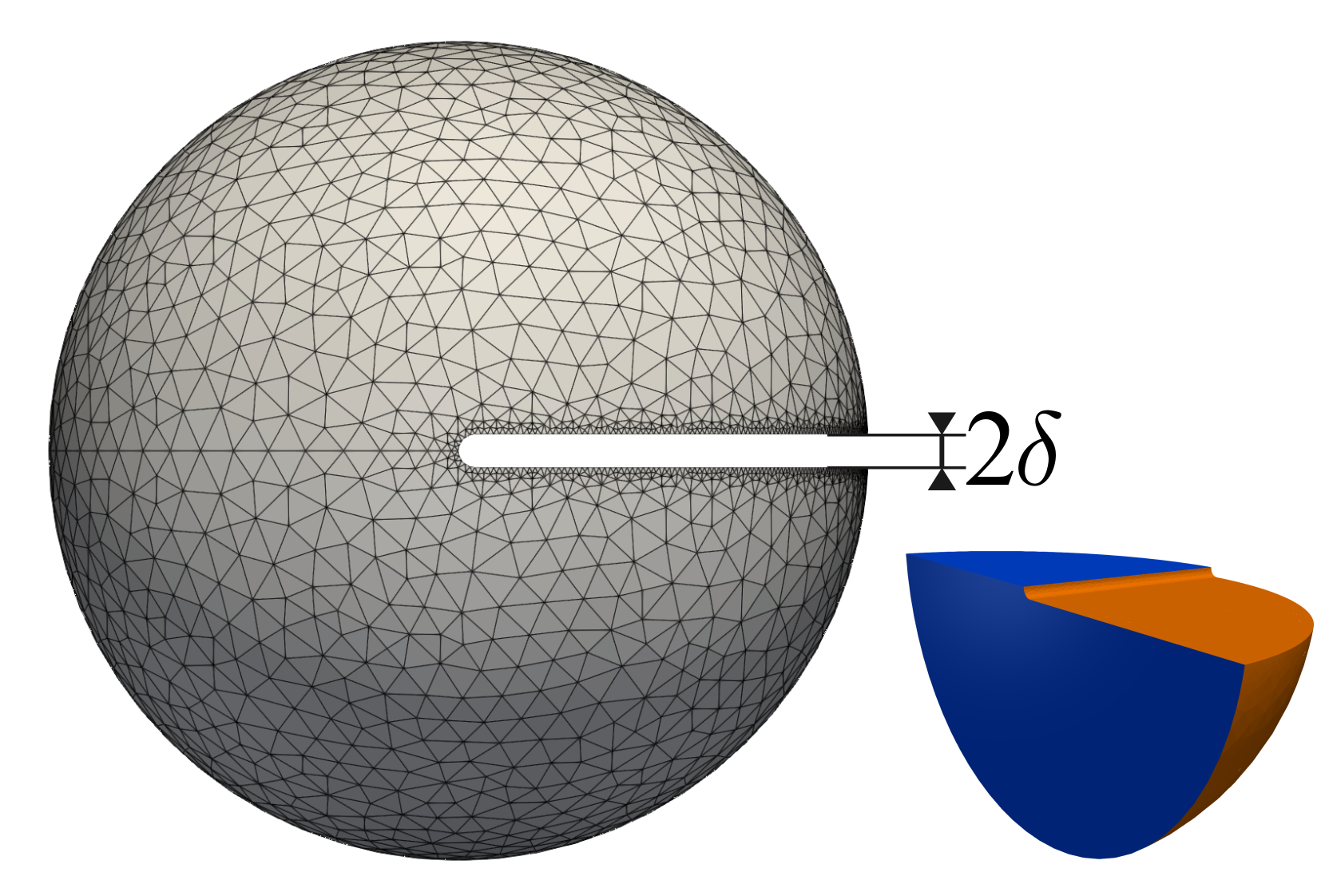}
  \caption{Representation of the full spheroid, with $\delta$ indicating the thickness of the cut. In the numerical simulations, only a quarter of the cut spheroid, as shown in the inset, is modelled. The blue surface in the inset corresponds to the region with symmetry boundary conditions, while the orange surface denotes the free boundary subject to surface tension. The mesh is refined near the cut edges to improve accuracy in these regions.}
  \label{fig:comp_domain}
\end{figure}

\begin{figure*}[t!]
  \includegraphics[width=\textwidth]{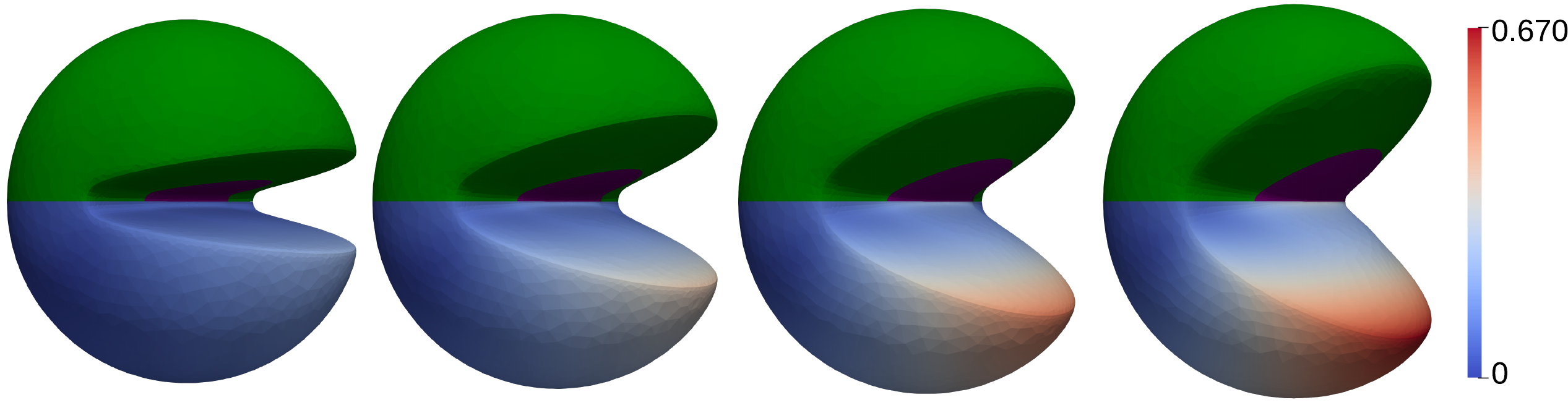}
  \caption{Current configuration of a cut spheroid after growth, with $R_T=2R_0$. From left to right, $\ell_c/R_T = 0.05,\,0.1,\,0.15,\,0.2$. The colour-bar denotes $\|\vect{u}\|/R_T$. The incision depth is $L = R_T$. On the top half of the spheroid, we denote in purple the initial region occupied by the spheroid at $t=0$, while the green part indicates the grown region.}
  \label{fig:3D}
\end{figure*}

We observe that for moderate values of the dimensionless parameter $\ell_c/ R_o$ there is a very good agreement between the analytical and the numerical model. When $\ell_c/R_0>1.5$ the discrepancy between the two models starts to be more relevant. This can be due to the non-homogeneity of the deformations close to the corners. In Fig.~\ref{fig:2D_map}, we present the current configurations of the cut sphere for various values of the incision length $L$ and of the ratio $\ell_c/R_0$. A quantitative plot of the different opening length observed as we change the parameters of the model is proposed in Fig.~\ref{fig:2D_data}, together with a comparison with the experimental results of Guillaume et al. \cite{Guillaume_2019} for a newly created spheroid (i.e. without growth). We observe that surface tension has a strong influence on the opening length (i.e. as we vary $\ell_c/R_0=\gamma/\mu R_0$). Conversely, the incision depth as a smaller influence on the resulting deformations. From the experimental data, we can estimate that $\ell_c/R_0$ is around $0.15$, even though there is some variance in the experimental results. Given that the shear modulus is approximately $1\,\mathrm{kPa}$ and the radius is around $450\,\mu\mathrm{m}$ \cite{Guillaume_2019}, the surface tension acting on the spheroid is estimated to be about $68 \,\mathrm{mN/m}$.

In the experiments reported in \cite{Guillaume_2019}, spheroids observed for growth have a reduced initial cell count. This has important consequences for the shape of the spheroid, as it retains a more spherical form. Consequently, we must abandon the assumption of cylindrical symmetry to model spheroid incision after growth, necessitating three-dimensional simulations to replicate these experiments.
\subsection{Simulations of three-dimensional spheroid cutting}

In \cite{Guillaume_2019} the authors provides data on the opening lengths observed after cutting spheroids that had been cultured for six days, reporting a doubling of their initial radius. Therefore, in what follows we focus on the case where the final radius is twice the initial radius, namely $R_T = 2 R_0$. These spheroids are characterized by an initial cell count of 500, with a final radius $R_T\sim450\mu\mathrm{m}$.

\subsubsection{Implementation}

The transition from 2D to 3D geometries introduces additional computational challenges. While the weak form remains consistent with that in~\eqref{eq:finite_element_pb}, the issue arises at the cut in the middle of the spheroid. Here, the cut spans the entire line, leading to a blow-up of the stress and pressure fields due to the infinite curvature at the tip of the cut \cite{Mora_2013,Mora_2015}. These singularities induce large deformations near the cut tip, which can create difficulties in the convergence of Newton's algorithm. To address this issue and model the mass depletion during the cutting process, we represent the cut as a crack with a finite thickness $2\delta$, as illustrated in Fig.~\ref{fig:comp_domain}. Specifically, in the following, we use a value of $0.05R_T$ for $\delta$.

\begin{figure}[t!]
  \includegraphics[width=\columnwidth]{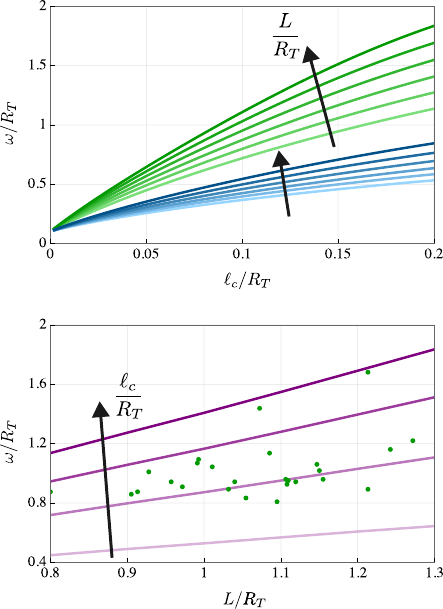}
  \caption{Plot of the opening $\omega$ of the cut sphere as a function of the elastocapillary length (top) and of the incision length (bottom). In the top panel, green curves indicate the opening of a spheroid after growth, in blue we show the results of a cut sphere without growth ($R_0=R_T$). All lengths are non-dimensionalised with respect to $R_T$. The value of $L/R_T$ in the top panel are $0.8,\,0.9,\,\dots,\,1.3$, while the values of $\ell_c/R_T$ in the bottom panel are $0.05,\,0.1,\,0.15,\,0.2$. In the bottom panel we also show the experimental data for a spheroid with $500$ cells at the beginning of the experiment and cut after six days, from \cite{Guillaume_2019}.}
  \label{fig:3D_data}
\end{figure}

\subsubsection{Results of the three-dimensional simulations}

In Fig.~\ref{fig:3D}, we show the current configuration of a grown spheroid following the cut for different values of $\ell_c/R_T$. We observe, as for the two-dimensional simulations, an increment in the tumour opening as the surface tension increases. The sharp edges of the cut are rounded by the effect of surface tension \cite{Mora_2015}.

The residual stresses accumulated in the spheroid during surface growth nearly double the tumour's opening compared to simulations without growth, as illustrated in Fig.\ref{fig:3D_data} (top). The simulation results without surface growth (blue lines) are comparable to those shown in Fig.\ref{fig:2D_data} for a circular domain, suggesting that the domain geometry plays a minor role in the mechanics of tumour opening post-incision.

A comparison with the experimental measurements of the opening length from Guillaume et al. \cite{Guillaume_2019} is shown in Fig.~\ref{fig:3D_data} (bottom). We observe a nearly linear relationship between the incision depth and the measured opening, although the slope of the curves is relatively small. This indicates that incision depth plays a minor role in the overall quantitative opening of the tumour. Instead, surface tension seems to have a more dominant influence. From the experimental results, we estimate the elastocapillary length to be between $0.1R_T$ and $0.15R_T$. Given that the shear modulus is approximately $1\,\mathrm{kPa}$ and the radius is around $450\,\mu\mathrm{m}$ \cite{Guillaume_2019}, the surface tension acting on the spheroid is estimated to be in the range of $45\,\mathrm{mN/m}<\gamma<68\,\mathrm{mN/m}$. This range aligns with the estimates reported by Riccobelli and Bevilacqua \cite{Riccobelli_2020}, who estimated $\gamma\sim 100 \,\mathrm{mN/m}$ using a different methodology.

\begin{figure}[t!]
  \centering
  \includegraphics[width=0.8\columnwidth]{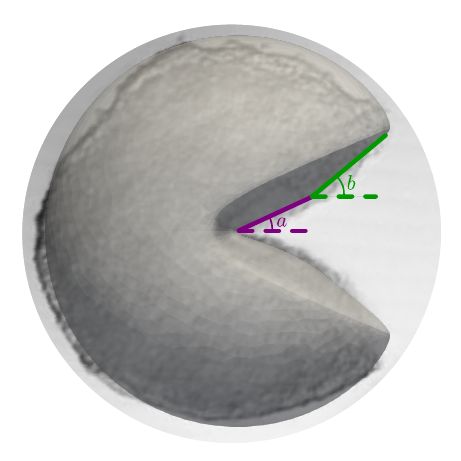}
  \caption{Superposition of the experimental image of a cut spheroid after 6 days of growth (initial number of cells: 500; image adapted from \cite{Guillaume_2019}) with a corresponding numerical simulation ($\ell_c/R_T=0.13$, $L/R_T=1.1$). The purple lines highlight the opening of the cut in the region already existing at $t=0$, showing an opening angle  of approximately $23^\circ$ (indicated with $a$ in the figure). The green lines represent the opening of the grown region of the spheroid, where the opening angle $b$ increases to around $40^\circ$.}
  \label{fig:3D_comparison}
\end{figure}

Interestingly, we observe that the residual stress also influences the shape of the cut: in the grown region, the opening of the tumour is more pronounced, as shown in Fig.~\ref{fig:3D_comparison}. This change in the opening is absent if the spheroid is cut just after its creation, see Fig~\ref{fig:browning} (right). In Fig.~\ref{fig:3D_comparison}, we remark that there is a strong agreement between the shape predicted by the numerical simulations and the one observed in the experiments.

\section{Discussion and concluding remarks}
\label{sec:conclusions}
Despite the large amount of literature on the mechanics of solid tumours, even basic aspects of cell proliferation and solid stress development are poorly understood. The theoretical framework of this paper provides a new perspective on these fundamental issues. As discussed in \cite{Riccobelli_2020}, we argue that surface tension plays a crucial role in determining the stress state and guiding cell proliferation in the early stages of tumour growth.

We have focused our attention on solid tumour spheroids, proposing a mathematical model of surface accretion based on the theory of Zurlo and Truskinovsky \cite{Zurlo_2017,Zurlo_2018,Truskinovsky_2019}. Surface tension elastically deforms newly produced cells on the free surface, resulting in a compression of the cells along the radial direction. This process generates an inhomogeneous prestretch and a geometrically incompatible relaxed state. The pattern of residual stress predicted by the model resembles that observed in real tumours  \cite{Stylianopoulos_2012,Ambrosi_2017}. The model predicts a structure of the spheroid composed of an inner core subject to isotropic compression and an outer rim where cells are radially compressed and circumferentially elongated, consistently with experimental observations, see Fig.~5 in \cite{Dolega_2017}.

To validate our theoretical framework, we have investigated the deformation of spheroids following radial incisions. In \cite{Guillaume_2019}, different opening length are observed depending on the spheroid's age. Tumour spheroids typically require two days to self-aggregate. After this period, the bulk of the spheroid shows negligible residual stress \cite{Lee_2019}. Instead, the primary source of stress within the spheroid can be attributed to the tissue surface tension, resulting in a pressure-like stress state in bulk \cite{Riccobelli_2020}. We have shown that the radial cut generates an opening of the spheroid driven by surface tension, even in the absence of residual stress. We derived a simplified analytical approximation using classical universal solutions of nonlinear elasticity, which provides an estimation of tumour opening. However, this model neglects the non-uniform deformations arising from singular stresses at the cut edges, which are caused by the infinite curvature of the free surface in the reference configuration. %

To %
overcome these limitations, we have performed finite element simulations. The results of these simulations are reported in Figs.~\ref{fig:th_num_2D}-\ref{fig:2D_data}. We have shown that, despite its simplicity, our analytical approximation is in good agreement with the outcomes of the numerical simulations. We have also included the effect of growth and the related mechanical stress in older spheroids. The results of the simulations for growing spheroids are reported in Figs.~\ref{fig:3D}-\ref{fig:3D_comparison}. The morphology predicted by the numerical simulations strikingly matches with the experimental evidence.
The numerical simulations in early and grown spheroids are compared with the quantitative data provided by Guillaume et al. \cite{Guillaume_2019}. Both cases indicate that the surface tension acting on the spheroids is about $60 \,\mathrm{mN/m}$, similarly to previous estimates \cite{Riccobelli_2020}.

The results of this study show the important role played by surface tension and elasticity in the growth of solid tumours, aligning with observations from several studies \cite{Yadav_2022,Yousafzai_2022,Ang_2024}. The proposed model sheds light on the influence of surface tension on tumour growth and elasticity, effectively capturing both the proliferation patterns within the tumour and the development of mechanical stress during the early stages of tumour progression. These fundamental mechanisms are crucial for understanding the subsequent processes of tumour vascularization and potential tissue invasion in cancer.

Open challenges related to this research are closely tied to mechano-transduction phenomena in solid tumours. Mechanical stress is known to inhibit growth, and while a mathematical framework for incorporating mechanical feedback into volumetric growth laws is well-established, its extension to surface growth remains non-trivial. Another key issue involves extending the proposed model to include compressible constitutive laws. Experimental evidence suggests distinct rheologies for the extracellular matrix and cells, with the former being compressible and the latter incompressible \cite{Dolega_2021}. In this respect, the extracellular matrix plays an important role in regulating cell proliferation and motility \cite{Dolega_2021_2,Ang_2024}.

While the current model captures the physical mechanisms underlying residual stress generation in tumour spheroids, a more refined model would require the extension to compressible nonlinear elastic and poroelastic materials \cite{Ambrosi_2017,dehghani2020role,Ang_2024}. However, spheroid compressibility can induce discrepancies between the reference and current configurations. This leads to a more complex reconstruction of the surface accretion process, as discussed in \cite{Truskinovsky_2019}. More generally, a coupled theory between surface and volumetric growth and remodelling would be important to accurately model biological tissues.

\subsubsection{Acknowledgments} I would like to express my gratitude for the insightful discussions with D. Ambrosi, G. Cappello, and G. Zurlo.

\subsubsection{Funding} This work has been partially supported by INdAM through the project \emph{MATH-FRAC: MATHematical modelling of FRACture in nonlinear elastic materials} and by PRIN 2022 project \emph{Mathematical models for viscoelastic biological matter}, Prot. 202249PF73 – Funded by European Union - Next Generation EU - Italian Recovery and Resilience Plan (PNRR) - M4C1\_CUP D53D23005610001. Financial support from the Italian Ministry of University and Research (MIUR) through the grant "Dipartimenti di Eccellenza 2023-2027 (Mathematics Area)” is gratefully acknowledged.

\subsubsection{Data availability statement} The source code and experimental data from \cite{Guillaume_2019} are available at \url{https://doi.org/10.5281/zenodo.14975959}

\printbibliography %

\end{document}